\documentclass[preprint,pre,aps]{revtex4}
\usepackage{graphics}
\usepackage{amsmath,amssymb}
\usepackage[dvips]{graphicx}
\begin{document}
\title{Spatial inhomogeneities  in ionic liquids, charged proteins and charge stabilized colloids from collective variables theory}
 \author{O. Patsahan}
\affiliation{Institute for Condensed Matter Physics of the National
Academy of Sciences of Ukraine, 1 Svientsitskii Str., 79011 Lviv, Ukraine}
 \author{ A. Ciach}
\affiliation{Institute of  Physical Chemistry, Polish  Academy of Sciences, 01-224 Warszawa,Poland}
 \date{\today}
\begin{abstract}

Effects of size and charge
asymmetry between  oppositely charged ions or particles on spatial inhomogeneities 
 are studied for a large range of charge and size ratios.
We perform a stability
analysis of the primitive model (PM) of ionic systems with respect to
 periodic ordering using the collective
variables based theory.  We extend previous studies  [A. Ciach et al., Phys. Rev.E \textbf{75}, 051505 (2007)]
in several ways.
First, we employ  a non-local  approximation for the reference
hard-sphere fluid which leads to the Percus-Yevick pair direct
correlation functions for the uniform case. Second, we use the
Weeks-Chandler-Anderson regularization scheme  for the Coulomb
potential inside the hard core. We  determine the relevant order
parameter connected with the periodic ordering and analyze the
character of the dominant fluctuations along the $\lambda$-lines. We
show that the above-mentioned modifications produce large quantitative and
partly qualitative changes in the  phase diagrams obtained
previously. We discuss possible scenarios of the periodic ordering
for the whole range of size- and charge ratios of the two ionic
species, covering electrolytes, ionic liquids, charged globular proteins or nanoparticles
 in aqueous solutions and charge-stabilized colloids.
\end{abstract}

\maketitle
\section{Introduction}
The study of phase diagrams of ionic systems in which the phase
separation is mainly driven by electrostatic forces is of great
fundamental interest and practical importance. Electrolyte
solutions, molten salts, ionic liquids and charge-colloidal
suspensions are examples of systems with dominant Coulomb
interactions. The  strong correlations between  ions and
counter-ions  are also known to play an important  role in
determining structure and phase behavior  of   micelles,
polyelectrolytes  and  proteins. Over the last few decades, the
phase behavior of ionic fluids has been the subject of many
experimental, theoretical  and simulation studies and the reviews of
the state of the art in this field   are  available in
Refs.~\cite{stell1,levinfisher,Schroer:review,Schroer:review:12,Panagiotopoulos:05,
Hynnien-Panagiotopoulos:08}. A great deal of the research  has been
focused on the fluid-fluid phase separation, whereas the fluid-solid
phase transition has received less attention so far.

On the other hand, recent experimental, simulation and theoretical
studies of room temperature ionic liquids (RTILs)
\cite{russina:11:0}, charged
 globular proteins~\cite{stradner:04:0,desfougeres:10:0} or colloidal particles
\cite{stradner:04:0,Arora:88,elmasri:12,Yamanaka:98,Royall:2004}
 reveal structural inhomogeneities of different type and spatial extent, and
it becomes evident that the ordering in systems dominated by Coulomb
interactions has a very rich and complex nature, and is far from
 being understood. Consequently, ordering of ions or charged particles on different length scales
attracts increasing attention.

The much studied  simple salts or electrolytes are quite well
understood. In these systems, the sizes of ions are similar.
Therefore, theoretical studies focused mainly on the restricted
primitive model (RPM) where ions of the same valence are modeled by
charged hard spheres of the same diameter immersed in a
structureless dielectric continuum. At large enough volume
fractions, transition to an ionic crystal occurs in molten salts
when temperature
decreases~\cite{stillinger,ciach-patsahan:1,Vega:96,Hynninen-06}. In
the ionic crystal the charge is periodically ordered and this
ordering is subjected to  the charge neutrality.  On the other hand,  charge-ordered,
neutral living clusters of various sizes and life times were found
in simulation studies of the fluid phase \cite{weis:98:0,spohr:02:0,panag,depablo}.

A direct inspection of the structure in ionic systems is complicated
due to a small size of the ions.  However, in compliance with  the
law of corresponding states one my expect a similar behavior in
oppositely charged colloidal particles of similar sizes under
appropriate rescaling of the length and energy units. The natural
length scale is
 the sum of radii of the anion and the cation, $\sigma_{\pm}$, while the energy unit is the Coulomb potential between
the oppositely charged ions at contact, $E_0=q_+q_-/(\epsilon \sigma_{\pm})$, where $q_{\alpha}$ is the charge of the
 $\alpha$-type ion ($\alpha=+,-$) and $\epsilon$ is the dielectric constant of the solvent. If the non-Coulomb
 interactions
 are negligible, the phase diagrams of systems consisting of spherical ions of various sizes  should have the same form in terms of the volume
 fraction and a
reduced temperature $T^*=k_BT/E_0$. Indeed, simulations show that
the phase diagram of charged particles with similar sizes in
deionized solvent (large Debye screening length)  resembles the
phase diagram of the
RPM~\cite{Hynninen-06,Caballero:07,Vega:review}. In the above
reduced units, the room temperature is very low  for simple salts,
whereas for colloidal particles it is high. Ordering of colloids at
room temperature
 is directly accessible to observations  and
 vacancies in the crystal can be  seen~\cite{Leunissen}.
Formation of neutral  aggregates  in the RPM agrees with recent
experimental studies of oppositely charged proteins of similar sizes
\cite{biesheuvel,desfougeres:10:0}. The aggregation was entirely
suppressed when the electrostatic interactions had been screened by
addition of a sufficient amount of salt. Thus, the electrostatic
interaction is  essential at least for the first step of the
self-assembly of oppositely
 charged globular  proteins  into aggregates. Large spherical aggregates presumably indicate nucleation of a crystal.

RTILs differ from simple salts mainly because the sum  of radii of
the ions is much larger. This  reduces the strength of the Coulomb
interactions at ion contact and, in turn, increases   the room
temperature in the reduced units. In addition, in many RTILs the
size and shape of the anion can be quite  different from the size
and shape of the cation. Moreover, non-Coulomb interactions may play
an important role here. In such a case the RPM is
 an oversimplification, and the above discussion does not apply to the RTILs.
Recent experimental and simulation studies on a series of
imidazolium-based RTILs indicate spatial inhomogeneities on the
length scale $\sim 10$~nm \cite{russina:11:0}. The inhomogeneous
structure of liquid resembles the structure of the
crystal~\cite{russina:11:0}. The origin of such nanoheterogeneity
has not been convincingly explained yet, and new studies are
required.

In a mixture of positively and negatively charged globular proteins,
the size difference between the positively and negatively
 charged molecules can be similar to the size difference between the anion and the cation in the RTILs. Recent experimental
studies show that the size difference plays an important role for the assembly of proteins into aggregates. According to
 Ref.~\cite{desfougeres:10:0}, formation of spherical aggregates  of oppositely charged proteins with
 overall charge near zero
requires charge- and size compensation.

The size difference between charged globular proteins and
counterions in a solution is much larger than in the previously
described cases. For example, the diameter of a lysozyme molecule is
of order of $3$~nm, i.e., it is 10 times larger than the counterions
in an aqueous solution. The charge of  small globular proteins is of
order of $10e$, where $e$ is the
 elementary charge. Scattering experiments~\cite{stradner:04:0} suggest clustering of protein molecules in water. Effective
 interactions between the lysozyme molecules were shown to have a form of short-range attraction and long-range
repulsion -- the latter resulting from the charges on the molecules
\cite{shukla}. The effective attraction  between two molecules is
caused by the interactions between them and the counterions
attracted to both molecules, as well as by the non-Coulomb forces.
Simulations  carried out in Ref.~\cite{kowalczyk} for the potential
derived in Ref.~\cite{shukla} show
 that the fraction of clustered molecules and the shape of clusters strongly depend  on the lysozyme
 volume fraction.

Charged nanoparticles or colloidal particles with a diameter $\sim
10\div 1000$~nm are several orders of magnitude larger than the
microscopic counterions in a solution, and their charge can be
orders of magnitude larger than the charge of the counterion. In
these systems, the experiments indicate the formation of colloidal
crystals with large interparticle distance for
 small volume fractions, and a re-entrant melting for larger volume fractions~\cite{Royall:2004,Arora:88,elmasri:12}.

From the above description of different experimental systems it follows that spatial inhomogeneities are common, but their
 extent and nature still need to be comprehended and classified for different sizes and charges of the anion and the cation.
 In the first step, it is essential to predict the structure formation and phase diagram
for various size and charge ratios for a generic model where
non-Coulomb interactions, deviations from a spherical shape  and
flexibility of  molecules are neglected. The role of the above
factors can be determined after comparing the results obtained
within a generic model  with the experiments.

The generic model that allows one to predict the phase separation
driven exclusively by Coulomb forces is a primitive model (PM). In
this model, the ionic fluid is described as an electroneutral
mixture of charged hard spheres immersed in a structureless
dielectric continuum.  The PM pair potential for two ions $\alpha$
and $\beta$ at distance $r$ apart is
\begin{eqnarray}
U_{\alpha\beta}(r) = \left\{
                     \begin{array}{ll}
                     \infty, & r<\sigma_{\alpha\beta}\\
                     \displaystyle\frac{q_{\alpha}q_{\beta}}{\epsilon r},&
                     r\geqslant \sigma_{\alpha\beta}
                     \end{array}
              \right. \,,
\label{int-PM}
\end{eqnarray}
where an ion of species $\alpha$ has a diameter $\sigma_{\alpha}$,
and charge $q_{\alpha}$,
$\sigma_{\alpha\beta}=\frac{1}{2}\left(\sigma_{\alpha}+\sigma_{\beta}\right)$
and $\epsilon$ is the dielectric constant. The PM is the simplest model for  all the above discussed systems.
The two-component PM
can be characterized by the parameters of  size and charge asymmetry:
\begin{equation}
\lambda=\frac{\sigma_{+}}{\sigma_{-}}\,, \qquad
Z=\frac{q_{+}}{|q_{-}|}\,. \label{model_par}
\end{equation}
For $\lambda=Z=1$ one arrives at the RPM.

In the PM with large difference in the ion radii, the spatial
distribution of the ions is expected to be quite different from  the
RPM, because the tendency for minimizing the electrostatic energy
competes with the geometrical
 restriction on packing of spheres with different sizes \cite{Leunissen}. Packing of large and small spheres that maximizes
the entropy could lead to the formation of mesoscopic charged
regions, and, on the other hand, when the electrostatic energy is
minimized, periodic pattern involving voids could be formed. As a
result of the competition between maximizing entropy and minimizing
energy, both the charge and the number-density of ions can oscillate
in space.

The  systematic studies of the   effects  of  size and charge
asymmetry on the periodic ordering  in  PMs  were initiated in
Ref.~\cite{Ciach-Gozdz-Stell-07} within the framework of the field
theoretical description. Based on a mean-field stability analysis,
the authors found the boundaries of stability of the disordered
phase for the whole range of $\lambda$ and $Z$. It was shown that,
besides a gas-liquid separation, in a certain portion of the phase
diagram, the uniform fluid became unstable with respect to the order
parameter oscillations of wavelengths $2\pi/k_{b}$ with $k_{b}\neq
0$. The line in the phase diagram corresponding to the instability
of the disordered phase with respect to periodic ordering is called
the $\lambda$-line~\cite{ciach:05} to distinguish it from the
spinodal line for which $k_b=0$.   The results   obtained in
Ref.~\cite{Ciach-Gozdz-Stell-07} show that (i) the periodic ordering
mainly depends on the size asymmetry; (ii) the qualitative
dependence on the charge asymmetry is found only for a sufficiently
large size asymmetry.

In this paper we continue the systematic study of the periodic
ordering in asymmetric PMs. We extend  the previous study in several
ways. The first modification concerns an approximate description of
the reference hard-sphere mixture.  In
Ref.~\cite{Ciach-Gozdz-Stell-07} a local-density approximation is
employed   for a hard-sphere free energy functional. Here, we
consider a non-local approximation for the reference hard-sphere
fluid which is shown to lead to the Percus-Yevick (PY)  theory for
the uniform case \cite{Rosenfield:89}. In our calculations we use
the Lebowitz's solution of the generalized PY  equation \cite{leb1}.

Another modification concerns  the regularization of the Coulomb
potential inside the hard core.  It is worth noting that in the
treatments of models with hard cores, the perturbation potential is
not defined uniquely inside the hard core. Here we  use the
Weeks-Chandler-Andersen  (WCA) regularization scheme  for the
Coulomb potentials  $\phi_{\alpha\beta}^{C}(r)$ \cite{wcha}
\begin{equation}
\phi_{\alpha\beta}^{C}(r) = \left\{
                     \begin{array}{ll}
                      \displaystyle\frac{q_{\alpha}q_{\beta}}{\epsilon \sigma_{\alpha\beta}},&
                     r< \sigma_{\alpha\beta} \\
                     \displaystyle\frac{q_{\alpha}q_{\beta}}{\epsilon r},&
                     r\geqslant \sigma_{\alpha\beta}.
                      \end{array}
              \right.
\label{WCA}
\end{equation}
instead of $\phi_{\alpha\beta}^{C}(r) =q_{\alpha}q_{\beta}\theta(r-\sigma_{\alpha\beta})/(\epsilon r)$ adopted in \cite{Ciach-Gozdz-Stell-07}.
As was shown in Ref.~\cite{cha}, the simple form for $\phi_{\alpha\beta}^{C}(r)$  given in Eq.~(\ref{WCA}) produces rapid
convergence of the series of the perturbation theory for the free
energy. On the other hand, the best theoretical estimates
for  the gas-liquid critical point of the RPM  was obtained using the WCA regularization scheme  \cite{patsahan_ion}.

Finally, we extend the study of the relevant order parameter (OP) undertaken in
\cite{Ciach-Gozdz-Stell-07}. Following the ideas of Refs.~\cite{Pat_physica,Patsahan-Patsahan},
we determine the  OP connected with the phase transition
to an ordered phase
and  analyze the character of the dominant fluctuations
along the $\lambda$-lines associated with the periodic ordering.

A theoretical background for this study is the statistical field
theory that exploits the method of collective variables (CVs)
\cite{zubar,jukh,Yukhn-book,Pat-Mryg-CM}. The theory enables us to
derive an exact expression for the functional of grand partition
function (GPF) of the model and on this basis to develop the perturbation
theory
\cite{Pat-Mryg-CM,patsahan-mryglod-patsahan:06,Patsahan_Patsahan:10}.
As was shown in Ref.\cite{patsahan-mryglod-patsahan:06},  the
well-known approximations for the free energy, in particular
Debye-H\"{u}ckel limiting law and the mean spherical approximation,
can be reproduced within the framework of this theory. Links between
this approach and the  field theoretical approach
\cite{Ciach-Gozdz-Stell-07} were established in
\cite{patsahan-mryglod:06} for the case of the RPM.

Our paper is organized as follows. In Section~2 we give some brief
background to the CVs based theory for the PM. Based on the Gaussian
approximation of the functional of GPF we obtain the pair direct
correlation functions and determine the OP characterizing the
periodic ordering in PMs. In Section~3 we study the effects of size
and charge asymmetry on the periodic ordering taking into account
the above-listed modifications. We discuss  the results  in
Section~4 and conclude in Section~5.

\section{Theoretical Background}
\subsection{Functional representation}

We start with the general case of a two-component PM consisting of
$N_{+}$ cations carrying a charge $q_{+}=Zq$ of diameter
$\sigma_{+}$ and $N_{-}$ anions carrying a charge $q_{-}=-q$ of
diameter $\sigma_{-}$. The ions are immersed in a structureless
dielectric continuum. The system is electrically neutral:
$\sum_{\alpha=+,-}q_{\alpha}\rho_{\alpha}=0$ and
$\rho_{\alpha}=N_{\alpha}/V$ is the number density of the $\alpha$th
species.

The pair interaction potential is assumed to be of the following
form:
\begin{equation}
U_{\alpha\beta}(r)=\phi_{\alpha\beta}^{\mathrm{HS}}(r)+\phi_{\alpha\beta}^{\mathrm{C}}(r),
\label{2.1}
\end{equation}
where $\phi_{\alpha\beta}^{\mathrm{HS}}(r)$ is the interaction
potential between the two  additive hard spheres of diameters
$\sigma_{\alpha}$ and $\sigma_{\beta}$. We call the two-component
hard-sphere system a reference system. Thermodynamic and structural
properties of the reference system are assumed to be known.
$\phi_{\alpha\beta}^{\mathrm{C}}(r)$ is the Coulomb potential given
in Eq.~(\ref{WCA}),  and hereafter we put $\epsilon=1$.

Using the CV method we get an exact functional representation of GPF for  the PM with size and charge
asymmetry \cite{Pat-Mryg-CM}
\begin{eqnarray}
\Xi[\nu_{\alpha}]&=&\int ({\rm d}\rho)({\rm d}\omega)\exp\left(
-\frac{\beta}{2V}\sum_{\alpha,\beta}\sum_{{\mathbf k}}\tilde
\phi_{\alpha\beta}^{C}(k)\rho_{{\mathbf k},\alpha}\rho_{-{\mathbf
k},\beta}+\right.\nonumber\\
&&\left. +{\rm i}\sum_{\alpha}\sum_{{\mathbf k}}\omega_{{\mathbf
k},\alpha}\rho_{{\mathbf k},\alpha}+\ln \Xi_{\rm{HS}}[\bar
\nu_{\alpha}-{\rm i}\omega_{\alpha}] \right).
\label{2.3}
\end{eqnarray}
In Eq. (\ref{2.3}) $\rho_{{\mathbf k},\alpha}=\rho_{{\mathbf
k},\alpha}^c-{\rm i}\rho_{{\mathbf k},\alpha}^s$   is the  CV which
describes the value of the $\mathbf k$-th fluctuation mode of the
number density of the $\alpha$th species, each of $\rho_{{\mathbf
k},\alpha}^{c}$ ($\rho_{{\mathbf k},\alpha}^{s}$) takes all the real
values from $-\infty$ to $+\infty$; $\omega_{{\mathbf k},\alpha}$ is
conjugate to the CV $\rho_{{\mathbf k},\alpha}$; $({\rm d}\rho)$ and
$({\rm d}\omega)$ are volume elements of the CV phase space
\begin{displaymath}
({\rm d}\rho)=\prod_{\alpha}{\rm d}\rho_{0,\alpha}{\prod_{\mathbf
k\not=0}}' {\rm d}\rho_{\mathbf k,\alpha}^{c}{\rm d}\rho_{\mathbf
k,\alpha}^{s}, \quad ({\rm d}\omega)=\prod_{\alpha}{\rm
d}\omega_{0,\alpha}{\prod_{\mathbf k\not=0}}' {\rm d}\omega_{\mathbf
k,\alpha}^{c}{\rm d}\omega_{\mathbf k,\alpha}^{s}
\end{displaymath}
and the product over ${\mathbf k}$ is performed in the upper
semi-space ($\rho_{-\mathbf k,\alpha}=\rho_{\mathbf k,\alpha}^{*}$, $\omega_{-\mathbf k,\alpha}=\omega_{\mathbf k,\alpha}^{*}$).

$\tilde
\phi_{\alpha\beta}^{\mathrm{C}}(k)$ is the Fourier transform of the
Coulomb potential. In the case of the WCA regularization (see Eq.~(\ref{WCA})) we obtain for $\beta\tilde \phi_{\alpha\beta}^{C}(k)$
\cite{Patsahan_Patsahan:10}:
\begin{eqnarray}
\beta\tilde\phi_{++}^{C}(k)&=&\frac{4\pi Z\sigma_{\pm}^{3}}{T^{*}(1+\delta)}\frac{\sin(x(1+\delta))}{x^{3}},
\label{Coulomb_WCA-1}
\\
\beta\tilde\phi_{--}^{C}(k)&=&\frac{4\pi\sigma_{\pm}^{3} }{T^{*}Z(1-\delta)}\frac{\sin(x(1-\delta))}{x^{3}},
\label{Coulomb_WCA-2}
\\
\beta\tilde\phi_{+-}^{C}(k)&=&-\frac{4\pi\sigma_{\pm}^{3} }{T^{*}}\frac{\sin(x)}{x^{3}},
\label{Coulomb_WCA}
\end{eqnarray}
where  the following notations are introduced:
\begin{equation}
T^{*}=\frac{k_{B}T}{E_{0}}=\frac{k_{B}T\sigma_{\pm}}{q^{2}Z}
\label{temp}
\end{equation}
is the dimensionless temperature, $x=k\sigma_{\pm}$,  $\sigma_{\pm}=(\sigma_{+}+\sigma_{-})/2$ and
\begin{equation}
\delta=\frac{\lambda-1}{\lambda+1}.
\label{delta}
\end{equation}
Similarly,    hereafter we introduce the parameter $\nu$
\begin{equation}
 \nu=\frac{Z-1}{Z+1}.
\label{nu}
\end{equation}
The parameters $\delta$ and $\nu$ are more convenient than the
parameters $\lambda$ and $Z$ because they vary between $-1$ and $1$. Following Ref.~\cite{Ciach-Gozdz-Stell-07}, we choose the
dimensionless temperature $T^{*}$ given in (\ref{temp}) and the
volume fraction of all ions
\begin{equation}
\zeta=\frac{\pi}{6}(\rho_{+}\sigma_{+}^{3}+\rho_{-}\sigma_{-}^{3})
\label{fraction}
\end{equation}
as thermodynamic variables.

$\Xi_{\rm{HS}}[\bar\nu_{\alpha}-{\rm i}\omega_{\alpha}]$ is the GPF
of a two-component hard-sphere system  with the renormalized
chemical potential
\begin{eqnarray}
\bar \nu_{\alpha}=\nu_{\alpha}+\frac{\beta}{2V}\sum_{{\mathbf
k}}\tilde\phi_{\alpha\alpha}^{C}(k)
\label{2.7}
\end{eqnarray}
in the presence of the local field $-{\rm i}\omega_{\alpha}(r)$. In (\ref{2.7}) $\nu_{\alpha}$ is the
dimensionless chemical potential,
$\nu_{\alpha}=\beta\mu_{\alpha}-3\ln\Lambda_{\alpha}$,
$\mu_{\alpha}$ is the chemical potential of the $\alpha$th species,
$\beta$ is the reciprocal temperature, $\Lambda_{\alpha}^{-1}=(2\pi
m_{\alpha}\beta^{-1}/h^{2})^{1/2}$ is the inverse de Broglie thermal
wavelength.

In order to develop the perturbation theory we present
$\ln\Xi_{\rm{HS}}[\bar\nu_{\alpha}-{\rm i}\omega_{\alpha}]$ in  the
form of the cumulant expansion
\begin{eqnarray}
\ln\Xi_{\rm{HS}}[\ldots]&=&\sum_{n\geq 0}\frac{(-{\rm
i})^{n}}{n!}\sum_{\alpha_{1},\ldots,\alpha_{n}}
\sum_{{\mathbf{k}}_{1},\ldots,{\mathbf{k}}_{n}}
{\mathfrak{M}}_{\alpha_{1}\ldots\alpha_{n}}(\bar\nu_{\alpha};
k_{1},\ldots,k_{n})\times\nonumber\\
&&\times
\omega_{{\bf{k}}_{1},\alpha_{1}}\ldots\omega_{{\bf{k}}_{n},\alpha_{n}}
\delta_{{\bf{k}}_{1}+\ldots +{\bf{k}}_{n}}, \label{2.11}
\end{eqnarray}
where $\delta_{{\bf{k}}_{1}+\ldots+{\bf{k}}_{n}}$ is the Kronecker
symbol. In  Eq. (\ref{2.11}) the $n$th cumulant
${\mathfrak{M}}_{\alpha_{1}\ldots\alpha_{n}}$ coincides with the
Fourier transform of the $n$-particle connected correlation function
of a two-component hard-sphere system \cite{Pat-Mryg-CM}.

\subsection{Gaussian approximation}

Having set  ${\mathfrak{M}}_{\alpha_{1}\ldots\alpha_{n}}\equiv 0$ for
$n\geq 3$, after  integration in Eq. (\ref{2.3}) over $\omega_{{\bf{k}},\alpha}$ one arrives at the  Gaussian approximation  for the functional of
GPF   \cite{Patsahan_Patsahan:10}
\begin{eqnarray}
\Xi_{{\text G}}[\nu_{\alpha}]=\Xi_{\rm{MF}}[\bar\nu_{\alpha}]\;\Xi'\int(\mathrm{d}\rho)
\exp\Big\{-\frac{1}{2V}\sum_{\alpha,\beta}\sum_{\bf
k}\tilde{\cal C}_{\alpha\beta}(k)\rho_{{\bf k},\alpha}\rho_{-{\bf
k},\beta}\Big\},
\label{Ksi-G}
\end{eqnarray}
where $\Xi_{\rm{MF}}$ is the GPF in the mean-field approximation and
$\Xi'=\displaystyle\prod_{\mathbf
k}\det[V^{2}{\mathfrak{M}}_{2}]^{-1/2}$ with ${\mathfrak{M}}_{2}$
being the matrix of elements ${\mathfrak{M}}_{\alpha\beta}(k)/V$.
$\tilde{\cal C}_{\alpha\beta}(k)$ is the Fourier transform of the
pair direct (vertex) correlation function in the random phase
approximation (RPA)
\begin{equation}
\tilde{\cal
C}_{\alpha\beta}(k)=\beta\tilde\phi_{\alpha\beta}^{C}(k)+
\tilde{\cal
C}_{\alpha\beta}^{\text HS}(k).
\label{C-alpha-beta}
\end{equation}
In (\ref{C-alpha-beta}), $\tilde{\cal C}_{\alpha\beta}^{\text
HS}(k)$ is the Fourier transform of the pair direct correlation
function of a two-component hard-sphere system. It is connected with
${\mathfrak{M}}_{\alpha\beta}(k)$ by the relation
$\tilde {\cal C}_{2}^{{\text
HS}}(k){\mathfrak{M}}_{2}(k)=\underline{1}$, where $\tilde {\cal
C}_{2}^{{\text HS}}(k)$ denotes the matrix of elements  $\tilde{\cal
C}_{\alpha\beta}^{\text HS}(k)$ and $\underline{1}$ is the unit
matrix. In the limit $k=0$, the $\tilde{\cal
C}_{\alpha\beta}^{{\text HS}}$ coincides with the coefficients
$a_{\alpha\beta}$ which are obtained in \cite{Ciach-Gozdz-Stell-07}
as a result of  the local-density approximation.

It is convenient to introduce CVs   which describe the fluctuation modes of
the total number  and charge density, $\rho_{{\mathbf k},N}$ and $\rho_{{\mathbf k},Q}$, by the relations
\begin{eqnarray}
\rho_{{\mathbf k},N}&=&\frac{1}{1+Z}\left(\rho_{{\mathbf k},+}+\rho_{{\mathbf
k},-}\right), \nonumber \\
\rho_{{\mathbf k},Q}&=&\frac{1}{1+Z}\left(Z\rho_{{\mathbf
k},+}-\rho_{{\mathbf k},-}\right).
\label{CV_new}
\end{eqnarray}
Then,  Eq.~(\ref{Ksi-G}) can be rewritten in terms of
$\rho_{{\mathbf k},N}$ and $\rho_{{\mathbf k},Q}$ as follows:
\begin{eqnarray}
\Xi_{{\text G}}[\nu_{\alpha}]=\Xi_{\rm{MF}}[\bar\nu_{\alpha}]\;\Xi'\int(\mathrm{d}\rho_{N})(\mathrm{d}\rho_{Q})
\exp\Big\{-\frac{1}{2V}\sum_{A,B}\sum_{\bf
k}\tilde{\cal C}_{AB}(k)\rho_{{\bf k},A}\rho_{-{\bf
k},B}\Big\},
\label{Ksi-G1}
\end{eqnarray}
where $A (B)=N,Q$ and
\begin{eqnarray}
\tilde{\cal C}_{NN}(k)&=&\frac{1}{(1+Z)^{2}}\left[\tilde{\cal C}_{++}(k)+Z^{2} \tilde{\cal C}_{--}(k)+2Z\tilde{\cal C}_{+-}(k)\right],
\nonumber \\
\tilde{\cal C}_{QQ}(k)&=&\frac{1}{(1+Z)^{2}}\left[\tilde{\cal C}_{++}(k)+\tilde{\cal C}_{--}(k)-2\tilde{\cal C}_{+-}(k)\right],
\nonumber \\
\tilde{\cal C}_{QN}(k)&=&\frac{1}{(1+Z)^{2}}\left[\tilde{\cal C}_{++}(k)-Z\tilde{\cal C}_{--}(k)+(Z-1)\tilde{\cal C}_{+-}(k)\right]
\label{C-A-B}
\end{eqnarray}
are the density-density, charge-charge and charge-density direct correlation functions, respectively.

In general, an equation for the boundary of stability  of the
uniform phase with respect to fluctuations   is
given by
\begin{equation}
\left.\det\,\tilde{\cal
 C}_{2}\right\vert_{k=k_{b}}=0,
 \label{spinodal}
\end{equation}
where $\tilde{\cal  C}_{2}$ denotes the matrix of elements
$\tilde{\cal C}_{AB}(k)$ (or $\tilde{\cal C}_{\alpha\beta}(k)$). The
corresponding wave vector $k_{b}$  is determined from the equation~\cite{ciach:05}
\begin{equation}
\label{k_b}
 \partial\det\,\tilde{\cal
 C}_{2}
/\partial k=0.
\end{equation}
The case $k=k_{b}=0$ corresponds to the gas-liquid-like separation
\cite{Ciach-Gozdz-Stell-07,Patsahan_Patsahan:10}. Here we are interested in the $\lambda$-line, the  boundary of stability  associated with  fluctuations of
the OP with $k=k_{b}\neq 0$. On the  $\lambda$-line,
the fluid becomes unstable with respect to the periodic ordering
indicating that there can be a phase transition to an ordered phase.

\subsection{Order parameter}

The determination of the  OP  is the important issue in the  phase transition theory of mixtures.
This problem has got a consistent and clear solution  within the given approach.

In order to determine  the  OP associated with the
periodic ordering we follow the ideas of
Refs.~\cite{Pat_physica,Patsahan-Patsahan}. First,  we diagonalize
the square form in Eq. (\ref{Ksi-G1}) by means of an orthogonal
transformation
\begin{eqnarray}
\xi_{{\mathbf k},1}&=t_{NN}\rho_{{\mathbf k},N}+t_{NQ}\rho_{{\mathbf k},Q}, \label{ksi1}\\
 \xi_{{\mathbf k},2}&=t_{QN}\rho_{{\mathbf
k},N}+t_{QQ}\rho_{{\mathbf k},Q}. \label{ksi2}
\end{eqnarray}
The explicit expression for coefficients   $t_{AB}$ are given in Appendix.
The corresponding eigenvalues $\varepsilon_{1}(k)$  and $\varepsilon_{2}(k)$ are found to be
\begin{equation}
\varepsilon_{1,2}(k)=\frac{1}{2}\left(\tilde{\cal C}_{NN}(k)+\tilde{\cal C}_{QQ}(k)\pm  \left[(\tilde{\cal C}_{NN}(k)-
\tilde{\cal C}_{QQ}(k))^{2}+4\tilde{\cal C}_{NQ}^{2}(k)\right]^{1/2}\right).
\label{epsilon-i}
\end{equation}
Based on the solutions of equations (\ref{spinodal})-(\ref{k_b}) we
find the eigenvalue which becomes equal to zero along the calculated
 $\lambda$-lines  for the fixed values of parameters $\delta$ and $\nu$.
We suggest that the corresponding eigenmode is connected with the relevant
OP.

 \begin{figure}[h]
 \centering
 \includegraphics[height=6cm]{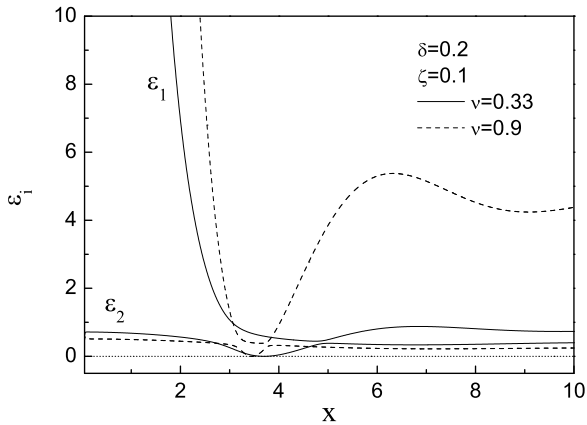}
 \caption{PM  with a small size asymmetry ($\delta=0.2$): the dependence of eigenvalues $\varepsilon_{1}$  and
$\varepsilon_{2}$  on the wave numbers ($x=k\sigma_{\pm}$).  Solid
and  dashed lines correspond to $\nu=0.33$ ($\zeta=0.1$,
$T^{*}\simeq  0.033 $) and $\nu=0.9$ ($\zeta=0.1$, $T^{*}\simeq
0.068$), respectively.     } \label{fig1}
 \end{figure}

Our analysis shows that  only one of the eigenvalues, i.e.,
$\varepsilon_{2}(k)$, becomes zero along the  $\lambda$-line of the
both symmetric and asymmetric PMs. The same result was obtained
earlier for a mixture of neutral particles
\cite{Pat_physica,Patsahan-Patsahan}.  In Figs.~1 and 2 we show a
typical behaviour of $\varepsilon_{1}(k)$ and $\varepsilon_{2}(k)$
under thermodynamic conditions corresponding to the points located
on the  $\lambda$-lines associated with $k_{b}\neq 0$.

We assume that CV $\xi_{{\mathbf k},2}$ is   connected to the relevant OP.
Based on Eqs.~(\ref{ksi1})-(\ref{ksi2}) we  can determine the direction
of strong fluctuations along the  $\lambda$-line by the relation
\begin{equation}
\label{theta}
\tan\theta=\frac{t_{NQ}}{t_{NN}}=-\frac{t_{QN}}{t_{QQ}},
\end{equation}
where  $\theta$ is the rotation angle of axes $\xi_{{\mathbf
k}_{b},1}$ and $\xi_{{\mathbf k}_{b},2}$ in the plane
($\rho_{{\mathbf k}_{b},N}$,$\rho_{{\mathbf k}_{b},Q}$). The case
$\theta=0$ corresponds to the pure charge density fluctuations (axes
$\xi_{{\mathbf k}_{b},2}$  and $\rho_{{\mathbf k}_{b},Q}$ coincide)
and the case $\theta=\mp\pi/2 $ corresponds to the pure total number
density fluctuations (axis $\xi_{{\mathbf k}_{b},2}$ coincides with
axes $\pm\rho_{{\mathbf k}_{b},N}$, respectively). Taking into
account the formulas from Appendix
(Eqs.~(\ref{tIJ})-(\ref{alpha_ij})), we have
\begin{equation}\label{theta_1}
\tan\theta=-\frac{1}{\alpha_{2}}=\alpha_{1  }.
\end{equation}
It is worth noting that in the long-wavelength limit
$\alpha_{2}(k=0)=0$, and one gets $\theta=-\pi/2$ in agreement with
the expected separation into homogeneous
 charge neutral dilute and dense phases.
 \begin{figure}[h]
 \centering
 \includegraphics[height=6cm]{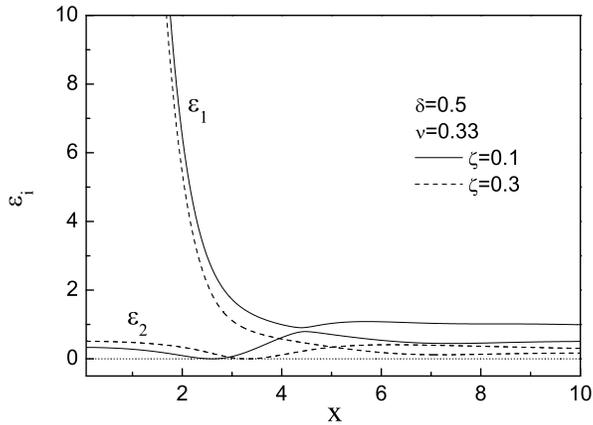}
 \caption{PM  with a moderate size asymmetry ($\delta=0.5$, $\nu=0.33$): the dependence of eigenvalues $\varepsilon_{1}$  and
$\varepsilon_{2}$  on the wave numbers ($x=k\sigma_{\pm}$).
 Solid and
the  dashed lines correspond to $\zeta=0.1$, $T^{*}\simeq 0.033$ and $\zeta=0.3$, $T^{*}\simeq 0.037$,
respectively.}
\label{fig2}
 \end{figure}

In Ref.~\cite{Ciach-Gozdz-Stell-07} the analysis was similar, except
that the eigenvalues denoted by $\tilde C_{\phi\phi}(k)$ and $\tilde
C_{\eta\eta}(k)$
were defined directly in terms of
 $\tilde C_{\alpha\beta}$ in such a
 way that $\varepsilon_2(k)=\tilde C_{\phi\phi}(k)$ when $\tilde C_{+-}(k)>0$ and
 $\varepsilon_2(k)=\tilde C_{\eta\eta}(k)$
when $\tilde C_{+-}(k)<0$. $\tilde C_{\phi\phi}(k)$ and $\tilde
C_{\eta\eta}(k)$ defined in Ref.~\cite{Ciach-Gozdz-Stell-07} reduce
to the direct correlation functions for the charge and for the
number density, respectively  for $Z=1$ (i.e. for the RPM), and the
corresponding eigenmodes reduce for $Z=1$ to the charge and the
number density waves. The advantage of the present approach is that
the critical mode is associated with the same eigenvalue
$\varepsilon_2(k)$ independently of $k$, $\zeta$ and $T$. It is also
more convenient to present the nature of the eigenmode in terms of
the angle $\theta$, rather than in terms of the parameter $R$
introduced in Ref.~\cite{Ciach-Gozdz-Stell-07} in a way analogous to
$\tan\theta$ in Eq.(\ref{theta}).

\section{Boundary of Stability  Associated with Periodic Ordering: Random  Phase Approximation}

In this section we study the effects of size and charge asymmetry on
the boundary of stability against the fluctuations with $k=k_{b}\neq
0$. To this end, we use Eqs.~(\ref{spinodal})-(\ref{k_b}) taking
into account Eqs.~(\ref{Coulomb_WCA-1})-(\ref{Coulomb_WCA}). In
order to determine  the character of the dominant fluctuations  we
use Eq.~(\ref{theta}).

We take into account the $k$-dependence of the direct correlation
functions of the reference system using an exact solution of the
generalized PY  equation obtained by Lebowitz
\cite{leb1}. The explicit expressions for the Fourier transforms of
the OZ  partial direct correlation functions of a two-component hard
sphere system are given in
Refs.~\cite{Ashcroft_Langreth:67,Ashcroft_Langreth:68}. They are too
cumbersome to be reproduced here.

As in Ref.~\cite{Ciach-Gozdz-Stell-07}, we distinguish the three
regimes in size asymmetry: small size asymmetry, moderate and large
size asymmetry, and very large size and charge asymmetry. Each
regime is characterized  by a typical behaviour  of the  $\lambda$-lines.
As we will see below, the $\delta$-ranges of these regimes slightly
differ when compared to \cite{Ciach-Gozdz-Stell-07}. We also dwell
briefly on the size-symmetric case.

\subsection{Size-symmetric PM}
We start with a size-symmetric PM  corresponding to $\delta=0$ (or
$\lambda=1$). In Fig.~3  the  $\lambda$-line  associated with the
wave vector $k_{b}\neq 0$ is displayed.  As is seen, it is a
straight line identical to that obtained for the RPM
\cite{ciach1,patsaha_mryglod-04}: $T^{*}(k=k_{b})=S_{\lambda}\zeta$.
For the WCA regularization, $S_{\lambda}\simeq 0.285$ and
$x_{b}=k_{b}\sigma_{\pm}=k_{b}\sigma\simeq 4.078$
\cite{patsaha_mryglod-04}.
\begin{figure}[h]
\centering
\includegraphics[height=6cm]{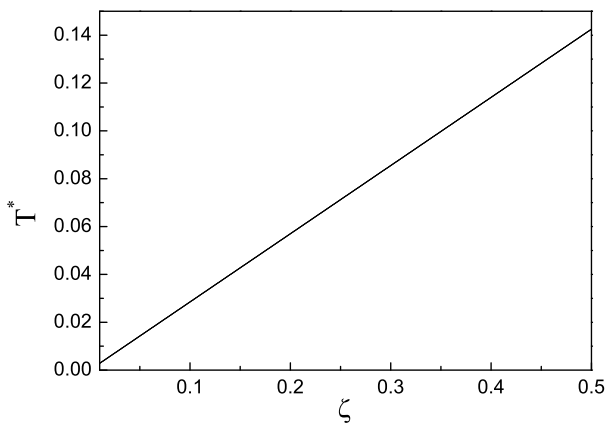}
\caption{The  $\lambda$-line for the transition to the ordered phase
for $\delta=0$. Temperature $T^{*}$ and the volume fraction of ions
$\zeta$ are in dimensional reduced units defined in
Eqs.~(\ref{temp}) and (\ref{fraction}), respectively.} \label{fig3}
\end{figure}

For the size-symmetric PM, the angle $\theta$ indicating the
direction of the strong fluctuations is equal to zero. Therefore,
the $\lambda$-line shown in Fig.~3 is the boundary of stability of
the uniform phase against the charge density fluctuations.
\begin{figure}[h]
\centering
\includegraphics[height=6.5cm]{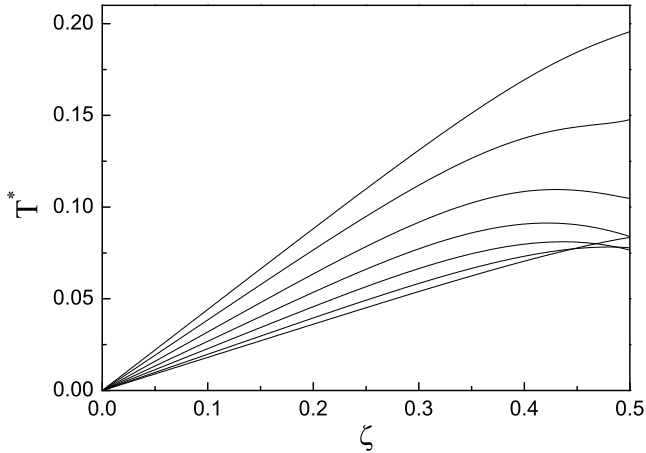}
\caption{The  $\lambda$-lines for the transition to the ordered
phase for $\delta=0.1$ ($\nu\geq 0$ and $\nu<0$). Lines from the top
to the bottom: $\nu=0.9$, $\nu=0.67$, $\nu=0.33$, $\nu=0$,
$\nu=-0.33$, $\nu=-0.67$, $\nu=-0.9$. Temperature $T^{*}$ and the
volume fraction of ions $\zeta$ are in dimensional reduced units
defined in Eqs.~(\ref{temp}) and (\ref{fraction}), respectively.}
\label{fig4}
\end{figure}

The charge asymmetry has no
effect on the  $\lambda$-line  of the size-symmetric PM. This
property does not persist if the higher-order terms  are taken into
account in Eq.~(\ref{2.11}) (see Refs.~\cite{Cai-JSP,Cai-Mol1,patsahan-mryglod-patsahan:06}). The
effects of charge-density fluctuations on a phase behaviour of RPM
($\delta=0$, $\nu=0$) were studied in Ref.~\cite{ciach-patsahan:1}
using the field theoretical description. It was found that in the
presence of fluctuations, the $\lambda$-line disappears. Instead, a
fluctuation-induced first-order transition to an ionic crystal
appears. Interestingly, the wavelength of the charge wave is
independent of the density along the  liquid-crystal coexistence
line \cite{ciach-patsahan:1}, which was interpreted as the formation
 of vacancies in the crystal of a fixed unit cell when the density of ions decreases.
We expect the similar situation to take place for the PM
with the small size asymmetry.

\subsection{Small size asymmetry}
Now we consider the  asymmetric PMs with a small size asymmetry
($\delta< 0.3$ or $\lambda< 2$).
This case corresponds to molten salts, electrolytes, some RTIL and to oppositely charged globular protein or nanoparticle
mixture in the limit of infinite screening length. The   $\lambda$-lines associated
with the wave vectors $k_{b}\neq 0$ are shown in Fig.~4 for
$\delta=0.1$. It should be noted that the  $\lambda$-lines  are
located at the temperatures that are  by an order of magnitude lower
than those in Ref.~\cite{Ciach-Gozdz-Stell-07}. In addition, the
monotonously increasing behaviour of the  $\lambda$-line
temperature, $T^{*}$ versus $\zeta$, is found for a  sufficiently
large charge asymmetry ($|\nu|>0.33$). For $|\nu|\leq 0.33$, $T^{*}$
has got a maximum  in the range  $\zeta\simeq 0.42\div 0,48$. The
effect comes into prominence with an increase of $\delta$. This
differs from the previous results \cite{Ciach-Gozdz-Stell-07}
demonstrating a monotonous increase of $T^{*}$ with  $\zeta$  along
the $\lambda$-lines  ($\zeta=0\div 0.7$) for all $\nu$. The
difference is directly related to the non-local approximation for
the reference hard-sphere system adopted in the present work. As is
seen from Fig.~4, the  $\lambda$-line temperature increases when
$\nu$ increases from $-0.9$ to $0.9$ which qualitatively agrees with
\cite{Ciach-Gozdz-Stell-07}.
\begin{figure}[h]
\centering
\includegraphics[height=6.5cm]{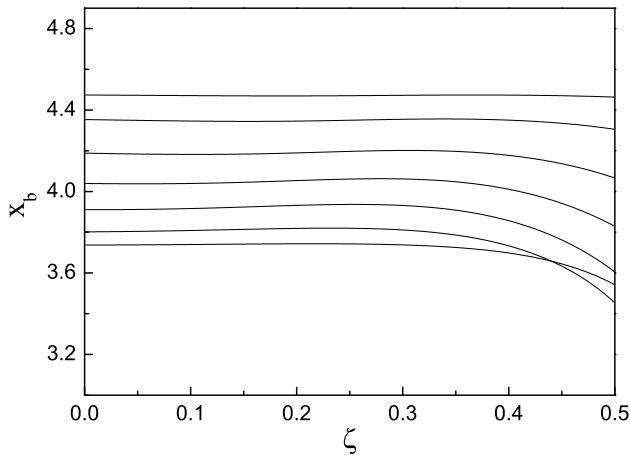}
\caption{The wave number $x_{b}=k_{b}\sigma_{\pm}$ corresponding to
the ordering of ions along the  $\lambda$-lines shown in Fig.~4 for
$\delta=0.1$ ($\nu\geq0$ and $\nu<0$). Lines from the bottom to the
top: $\nu=0.9$, $\nu=0.67$, $\nu=0.33$, $\nu=0$, $\nu=-0.33$,
$\nu=-0.67$, $\nu=-0.9$.  $\zeta$ is the volume fraction of ions.}
\label{fig5}
\end{figure}

For a small size asymmetry, the wave numbers characterizing the
period of the OP oscillations  in a non-uniform
phase are $x_{b}=k_{b}\sigma_{\pm}>\pi$ and their magnitudes depend
very slightly on $\zeta$ (see Fig.~5). The comparison with
Ref.~\cite{Ciach-Gozdz-Stell-07} implies  that the magnitude  of
$x_{b}$ is mainly determined by the regularization method of the
Coulomb potential inside the hard core. In Fig.~5 we also
demonstrate the effect of charge asymmetry on $x_{b}$. As is seen,
$x_{b}$ increases with the variation of $\nu$ from $0.9$ to $-0.9$
for the fixed  $\zeta$ which agrees with the results obtained in
\cite{Ciach-Gozdz-Stell-07}.
  \begin{figure}[h]
 \centering
 \includegraphics[height=6cm]{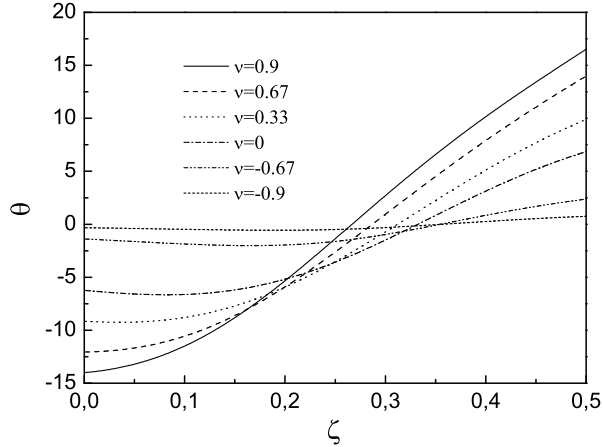}
 \caption{The variation of the angle $\theta$ along the  $\lambda$-lines  presented in Fig.~4 ($\nu\geq 0$ and $\nu<0$).
 $\theta$ is
measured in degrees.} 
\label{fig6}
 \end{figure}

We have calculated the angle $\theta$  showing the direction of strong
fluctuations along the  $\lambda$-line. The results for $\delta=0.1$
are presented in Fig.~6.   As is seen, $\theta$ changes around zero
and this change increases with an increase of size asymmetry. As
regards the charge asymmetry, the modulus of $\theta$ decreases with
the variation of  the charge asymmetry parameter  $\nu$ from $0.9$
to $-0.9$ for the fixed  $\delta$, and $\lvert \theta\rvert$
approaches zero for $\nu=-0.9$. In particular, for $\delta=0.1$ the
angle $\theta$ changes continuously in the range from $-14^{\circ}$
to $+16.5^{\circ}$  for  $\nu=0.9$ and  from $-0.3^{\circ}$ to
$+0.75^{\circ}$ for $\nu=-0.9$ when $\zeta$ is varied from $0$ to
$0.5$. For $\delta=0.2$ we have  $-30^{\circ}\lesssim\theta\lesssim
+44^{\circ}$ ($\nu=0.9$) and $-0.03^{\circ}\lesssim\theta\lesssim
+1.3^{\circ}$ ($\nu=-0.9$) in the same range of $\zeta$. In general,
the charge density fluctuations  are  the dominant fluctuations for
$\delta<0.3$. It should be noted that for a small size asymmetry,
the $\lambda$-lines associated with the periodic ordering are
located at much higher temperatures than the spinodals indicating
phase separation into two uniform phases.

Summarizing, in the  PM with a small size asymmetry  we  expect the
phase transition  to an ionic crystal with a compact unit cell where
nearest neighbours are oppositely charged,
 at least when $\zeta$ is
sufficiently large. In particular, the ${\rm CsCl}$ crystal was
observed experimentally for the system of oppositely charged
colloids of comparable sizes  \cite{Leunissen}. For low volume
fraction compact charge-ordered clusters are expected, in agreement
with experiments for oppositely charged proteins
\cite{desfougeres:10:0} and with simulations \cite{biesheuvel}.

\subsection{Moderate  and large size asymmetry}

Let us consider the case of moderate and large size asymmetry
corresponding to $\delta>~0.3$ ($\lambda>~2$).  In fact, PMs with
$\delta=0.3$ demonstrate a crossover-type behaviour. The
$\lambda$-lines  and the corresponding wave numbers for $\delta=0.3$
are shown in Figs.~7 and 8, respectively. It follows from our
calculations that PMs with $\delta=0.3$ and $\nu< -0.67$ do not
undergo an instability with respect to fluctuations with $k_{b}\neq
0$. This is  contrary to the results obtained for PMs with a small
size asymmetry (see Fig.~4). Similar to the small size asymmetry
case, the $\lambda$-line temperature decreases when $\nu$ varies
from $0.9$ to $-0.67$. For $\nu=-0.67$,   the $\lambda$-line
associated with $k_{b}\neq 0$ is located at $T^*$ slightly lower
than $T^*$ at the spinodal indicating the separation in two uniform
phases for the same $\zeta$.
\begin{figure}[h]
\centering
\includegraphics[height=6cm]{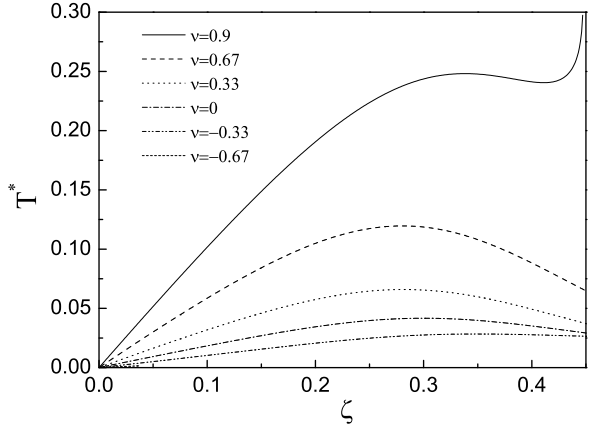}
\caption{The  $\lambda$-lines for the transition to the ordered
phase for $\delta=0.3$ ($\nu\geq0$ and $\nu<0$). Temperature $T^{*}$
and the volume fraction of ions $\zeta$ are in dimensional reduced
units defined in Eqs.~(\ref{temp}) and (\ref{fraction}),
respectively. The curve for $\nu=-0.67$ is located at a very low
temperature and is indistinguishable in this plot.} \label{fig7}
\end{figure}

As is seen from Fig.~8, the dependence of the wave numbers $x_{b}$
on the volume fraction $\zeta$ is more prominent than that shown in
Fig.~5. In particular, $x_{b}$ is an increasing function of $\zeta$
for  $\lvert \nu\rvert\leq 0.33$. Such behavior is consistent with the decrease of the interparticle separation
for decreasing average volume per particle for the same type of structure. For $\nu\geq 0.67$, $x_{b}$ first
very slowly increases   and then again slowly decreases  when volume
fraction increases. For $\nu= -0.67$, $x_{b}$ rapidly decreases. For
very small values of $\zeta$, $x_{b}\approx\pi$ ($l_{b}\approx\sigma_{+}+\sigma_{-}$)
without regard to the charge asymmetry.
\begin{figure}[h]
\centering
\includegraphics[height=6cm]{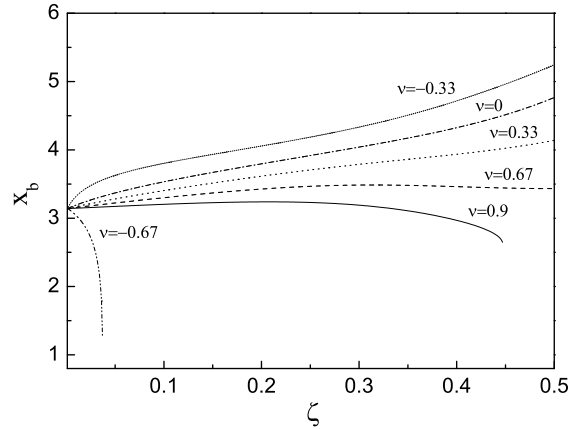}
\caption{The wave number $x_{b}=k_{b}\sigma_{\pm}$ corresponding to
the ordering of ions along the  $\lambda$-lines shown in Fig.~7 for
$\delta=0.3$.  $\zeta$ is the volume fraction of ions.} \label{fig8}
\end{figure}+

The variation of the direction of strong fluctuations
along the  $\lambda$-lines (angle $\theta$) is shown in Fig.~9. As is
seen, the charge density fluctuations still prevail over the total
number density fluctuations for $-0.33\leq \nu\leq 0.9$ except for a
high-density region (see the case $\nu=0.9$).

The fluid-solid phase coexistence in the PM with $\delta=0.3$ and
$\nu=0$ was studied by computer simulations in
Ref.~\cite{Hynnien-Panagiotopoulos:09}. In particular, it was found
that there is a coexistence at $T^{*}=0.033$  between  a fluid  at
$\zeta=0.38$ and a ($\rm{NaCl}$) solid phase at $0.51$. Our
calculations of the $\lambda$-line temperature corresponding to
$\zeta=0.38$ yield $T^{*}=0.037$ (see Fig.~7).   At this point, the
order parameter is connected to the charge density fluctuations
($\theta\approx 0^{\circ}$) suggesting the phase transition to the
ionic crystal.

For $\delta>0.3$ all the  $\lambda$-lines associated with $k_{b}\neq
0$ demonstrate a single maximum for the volume fraction $\zeta_{m}$
(see Figs.~10, 11  and 14). This implies that the periodic ordering is
less favorable at higher volume fractions. This behaviour is similar
to that found in \cite{Ciach-Gozdz-Stell-07} but for the higher
values of $\delta$, namely for $\delta>0.4$.   It is worth noting
that the  $\lambda$-lines  lie at temperatures lower than those
found in \cite{Ciach-Gozdz-Stell-07}. This tendency is kept for the
whole region of the variation of $\delta$. On the other hand, the
comparative analysis performed within the same regularization
scheme, i.e., the WCA regularization, shows that the account of the
$k$-dependence of the reference system correlation functions leads
to a pronounced increase of the  $\lambda$-lines temperatures for a moderate
and large size asymmetry (see Fig.~10). Such strong impact of the
above-mentioned $k$-dependence  on the magnitude of the
$\lambda$-lines temperature is  contrary to the results found for a
one-component fluid with competing attractive and repulsive
interactions \cite{Archer:08}.

  \begin{figure}[h]
 \centering
 \includegraphics[height=6cm]{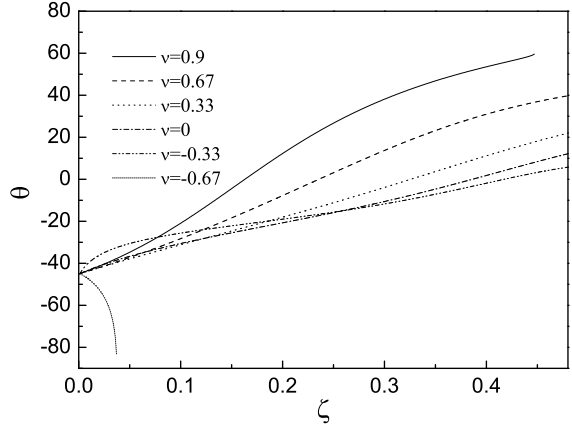}
 \caption{The variation of angle $\theta$ along the  $\lambda$-lines  presented in Fig.~7. $\theta$ is
measured in degrees.} 
\label{fig9}
 \end{figure}
\begin{figure}[h]
\centering
\includegraphics[height=6cm]{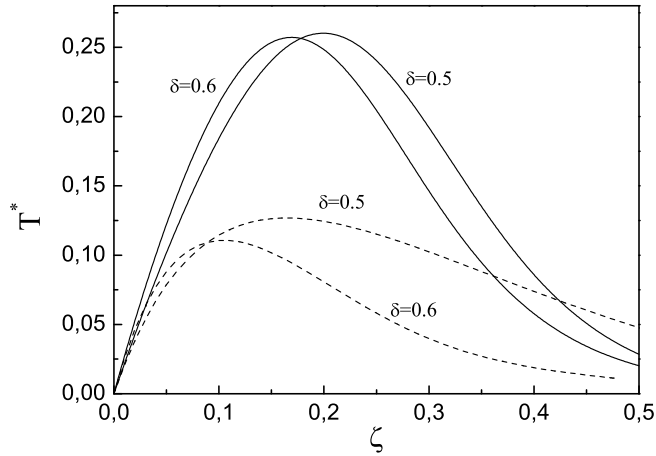}
\caption{The  $\lambda$-lines for the transition to the ordered
phase for   $\nu=0.9$ and  for two values of the size asymmetry
($\delta=0.5$ and $\delta=0.6$). Solid and dashed lines
correspond to a non-local and local approximations for the
hard-sphere reference system, respectively. Temperature $T^{*}$ and
the volume fraction of ions $\zeta$ are in dimensional reduced units
defined in Eqs.~(\ref{temp}) and (\ref{fraction}), respectively.}
\label{fig12}
\end{figure}
Unlike Ref.~\cite{Ciach-Gozdz-Stell-07},  our results demonstrate
the dependence of $\zeta_{m}$  on the both parameters $\delta$ and
$\nu$. In general, $\zeta_{m}$ decreases noticeably with an increase
of $\delta$ and has a non-monotonous behavior with the variation of
$\nu$. By contrast, the maximal value of the  $\lambda$-lines
temperature slightly depends  on the size asymmetry for $\delta>
0.3$.

Let us describe in some detail the case  $\delta=0.5$ ($\lambda=3$,
volume ratio $\sim 30$), characterizing moderate size asymmetry.
$\delta=0.5$ can be found in a solution of charged molecules of a
diameter $\sim 1$~nm in a presence of counterions of a diameter
$\sim 0.3$~nm, or in some RTIL  and finally in a mixture of
oppositely charged globular proteins or nanoparticles in deionized
solvent.
\begin{figure}[h]
\centering
\includegraphics[height=6cm]{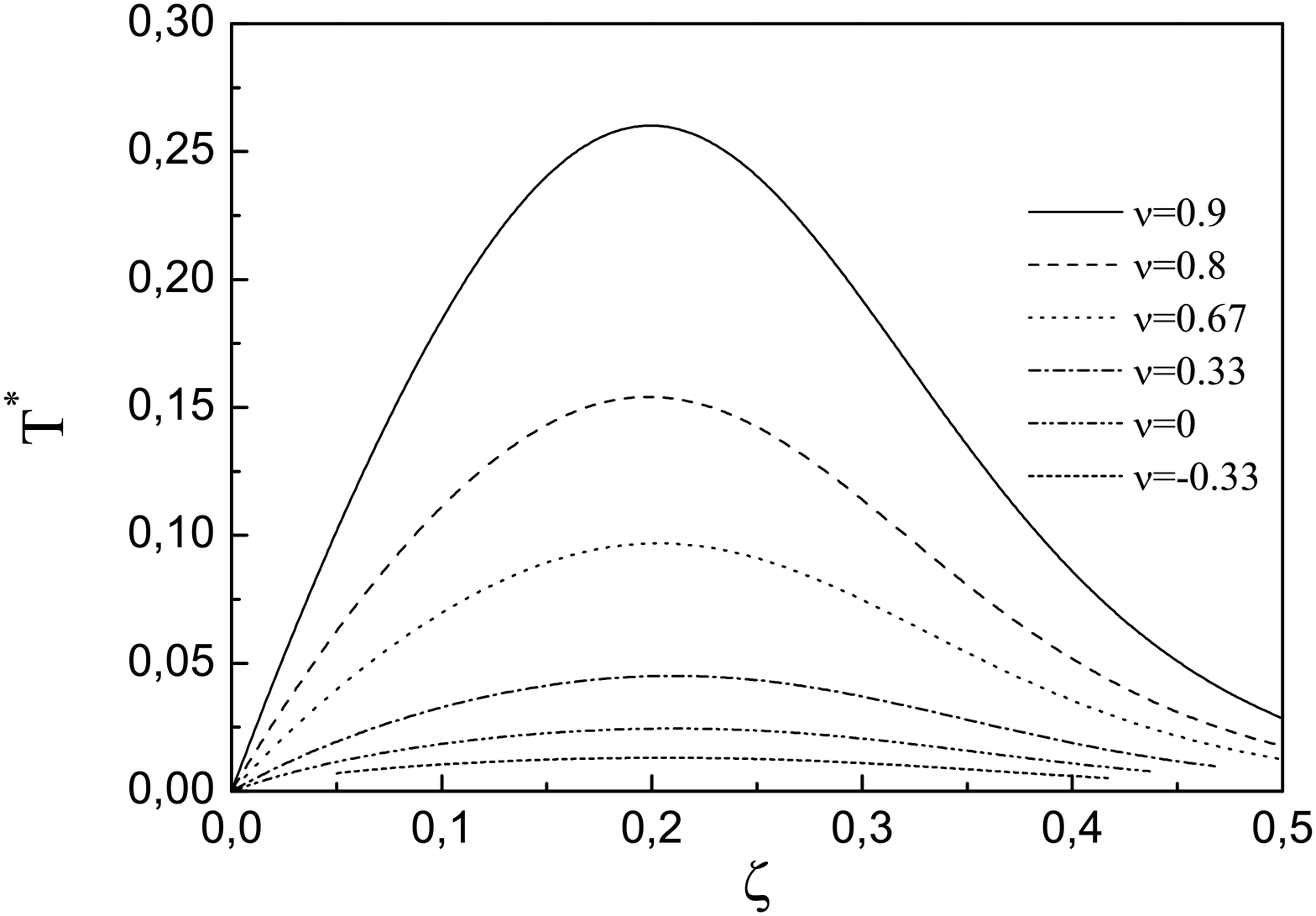}
\caption{The $\lambda$-lines for the transition to the ordered phase
for $\delta=0.5$. Temperature $T^{*}$ and the volume fraction of
ions $\zeta$ are in dimensional reduced units defined in
Eqs.~(\ref{temp}) and (\ref{fraction}), respectively.}
\label{fig110}
\end{figure}

For $\delta=0.5$,  the wave number $x_{b}$ increases along the
$\lambda$-line from $x_{b}<\pi$ for $\zeta\leq 0.2$ to $x_{b}>\pi$
for $\zeta\geq 0.2$ except for the case of large charge asymmetry
$\nu=0.9$  (see Fig.~12). Such a behaviour implies that different
ordered structures  can be  formed with the variation of $\zeta$.
For $\nu=0.9$, the wave number $x_{b}<\pi$  along the  $\lambda$-line
($0\leq\zeta\leq 0.5$) and depends only slightly on the volume
fraction.  For  $|\nu|<0.33$, $x_{b}$ change their trend sharply for
the large values of $\zeta$ ($\zeta\simeq 0.4$).
\begin{figure}[h]
\centering
\includegraphics[height=6cm]{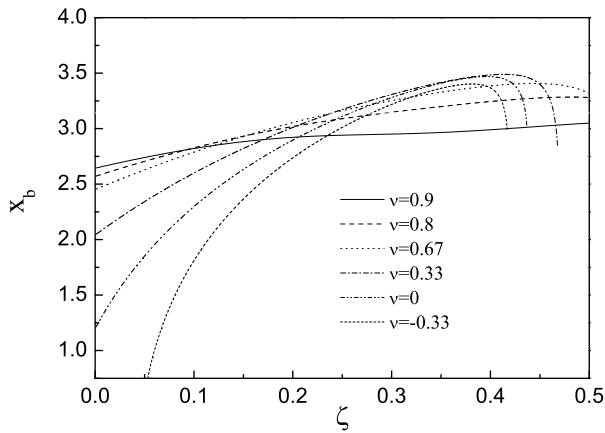}
\caption{The wave number $x_{b}=k_{b}\sigma_{\pm}$ corresponding to
the ordering of ions along the  $\lambda$-lines shown in Fig.~11 for
$\delta=0.5$.  $\zeta$ is the volume fraction of ions.}
\label{fig13}
\end{figure}
  \begin{figure}[h]
 \centering
 \includegraphics[height=6cm]{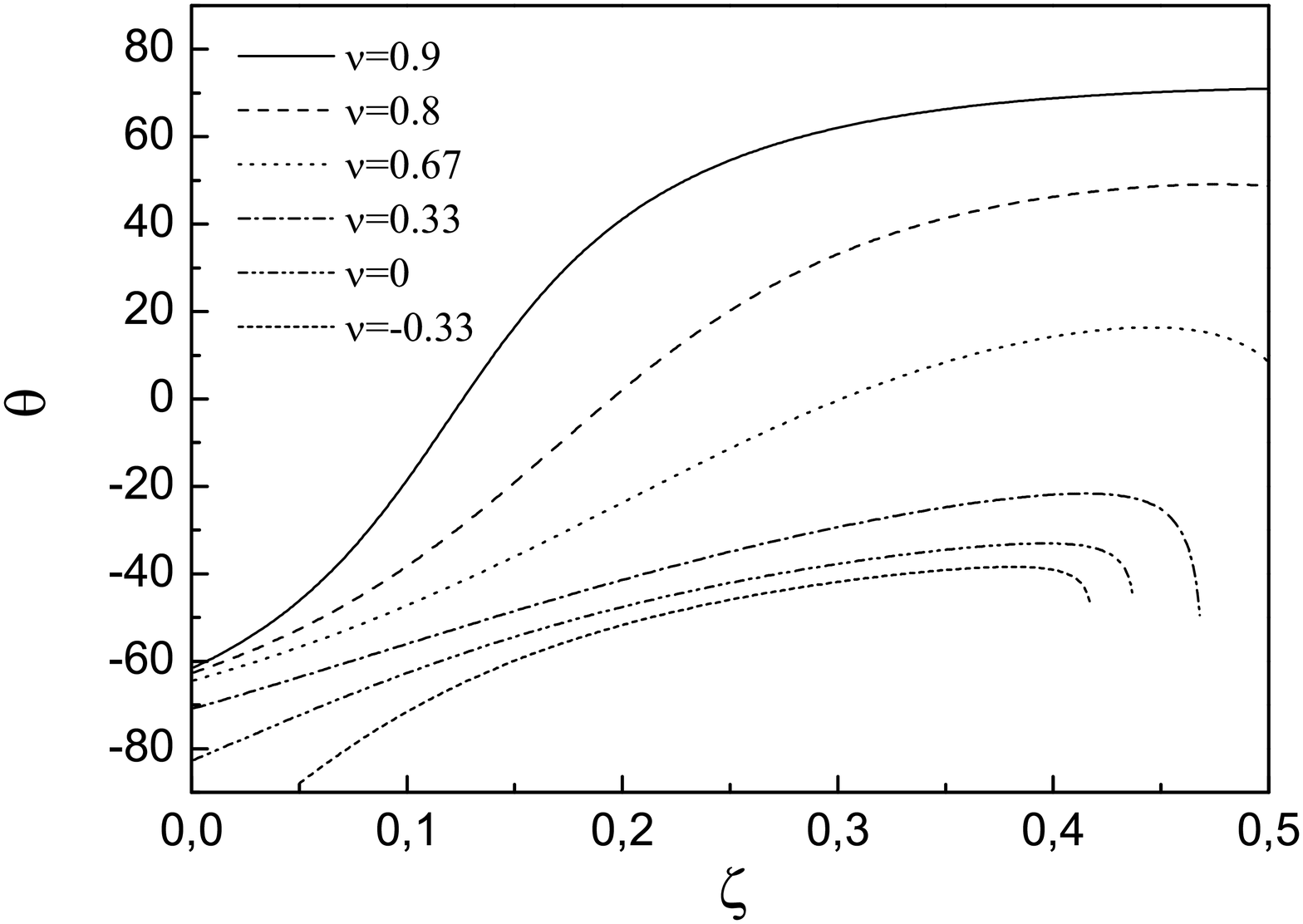}
 \caption{The variation of angle $\theta$ along the $\lambda$-lines  presented in Fig.~11. $\theta$ is
measured in degrees.} 
\label{fig15}
 \end{figure}
The dominant field is $\rho_{\mathbf{k},N}$  at
small $\zeta$ for all $\nu$ (see Fig.~15). For $\nu\geq 0.33$, the
character of the dominant fluctuations  changes with an increase of
$\zeta$. In particular, $\theta$ varies from $-61^{\circ}$ to
$+71^{\circ}$ for $\nu=0.9$ and from $-62^{\circ}$ to $+49^{\circ}$
for $\nu=0.8$ when the volume fraction increases from $0$ to $0.5$.
For $\nu=0.67$, the dominant field is $\rho_{\mathbf{k},Q}$ when
$\zeta\geq 0.11$. For $|\nu|\leq 0.33$, the angle $\theta$ takes a
maximum value $\theta_{m}$ for $\zeta\simeq 0.4$ and the modulus of
$\theta_{m}$ increases when $\nu$ varies from $0.33$ to $-0.33$:
$\mod\theta_{m}\simeq 22^{\circ}$ for $\nu= 0.33$ and
$\mod\theta_{m}\simeq 38^{\circ}$ for $\nu= -0.33$. Thus,  the
charge density fluctuations prevail over the total number density
fluctuations along the  $\lambda$-lines for this range of $\nu$.
In general, for the intermediate size asymmetry both the charge density and the total number density are inhomogeneous in space.
\begin{figure}[h]
\centering
\includegraphics[height=6cm]{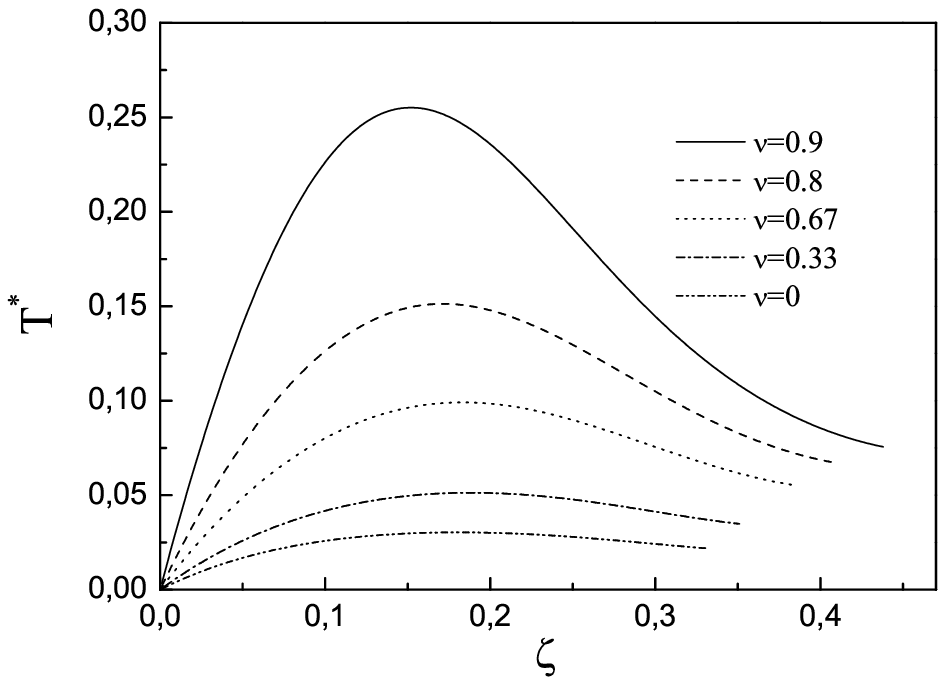}
\caption{The  $\lambda$-lines for the transition to the ordered
phase for $\delta=0.8$. Temperature $T^{*}$ and the volume fraction
of ions $\zeta$ are in dimensional reduced units defined in
Eqs.~(\ref{temp}) and (\ref{fraction}), respectively.} \label{fig11}
\end{figure}
\begin{figure}[h]
\centering
\includegraphics[height=6cm]{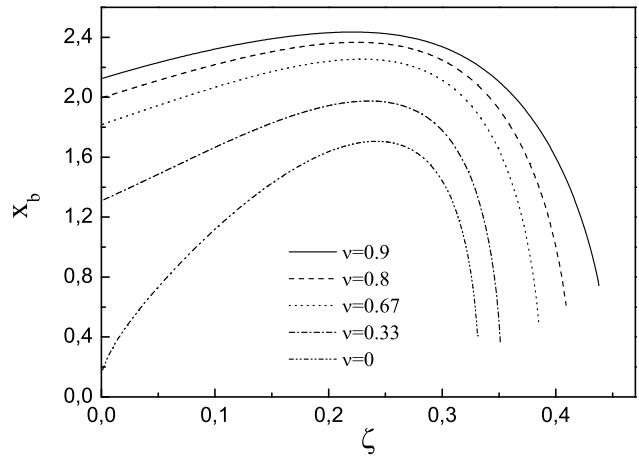}
\caption{The wave number $x_{b}=k_{b}\sigma_{\pm}$ corresponding to
the ordering of ions along the  $\lambda$-lines shown in Fig.~11 for
$\delta=0.8$.  $\zeta$ is the volume fraction of ions.}
\label{fig14}
\end{figure}

Let us focus on the large size asymmetry,  $\delta> 0.6$
($\lambda>4$, volume ratio $>60$). For $\delta\geq 0.6$,  we obtain
$x_{b}<\pi$ in the region $0\leq \zeta\leq 0.5$. The $\lambda$-line and the dependence of
$x_{b}$ on the volume fraction for $\delta=0.8$ are shown in Figs.~14 and 15.
As is seen,  $x_{b}$ first increases reaching its maximal value and
then it decreases with the increase of $\zeta$. The maximum becomes
more  pronounced with the decrease of charge asymmetry. Such  a
behaviour of $x_{b}$ is typical of the systems with
$0.6\leq\delta\leq 0.9$. The variation of the angle $\theta$ along
the  $\lambda$-lines for $\delta=0.8$ is shown in Fig.~16.  For the
charge asymmetry range $0\leq\nu\leq 0.8$, the total number density
fluctuations are the dominant fluctuations along the
$\lambda$-lines. Moreover, the contribution from this type of
fluctuations increases when the charge asymmetry decreases. For
$\nu=0.9$, the character of the dominant fluctuations changes
continuously along the  $\lambda$-line from the total number density
fluctuations for $0\lesssim\zeta\lesssim 0.12$ to the charge density
fluctuations  reaching a maximal value at $\xi\simeq 0.2$. For
$\zeta\gtrsim 0.28$, the dominant fluctuations are again the total
density fluctuations. It should be noted that for $\delta=0.9$ the
total density fluctuations play the  dominant role along  the whole
$\lambda$-line for $0\leq\nu\leq 0.9$.

In the
approximation considered, PMs with $\delta> 0.6$ and $\nu< 0$ do not
undergo instabilities against the fluctuations with $k_{b}\neq 0$.
Moreover, the  $\lambda$-lines  tend to
lower $T^*$ with the decrease of the charge asymmetry coming close
to the spinodals associated with the phase separation into uniform
phases. This implies that the phase transition to the periodic
ordering in the PMs with large size asymmetry becomes less favorable
when  the charge asymmetry decreases.

In conclusion, for the PMs with moderate and large  size   and
charge asymmetry, we expect the phase transition to the colloid
crystals of different structure formed by a nearly charge-neutral
units for small and large volume fractions. The size of these units
depends on the charge ratio and is generally larger than in the case
of $\delta<0.3$. For volume fractions near the maximum temperature
at the $\lambda$-line the number density in positively charged
regions is different from the number density in negatively charged
regions in the fluctuation that destabilizes the homogeneous phase
(see Figs.~14 and 16). Global stability, however, may correspond to
another structure. For  a large size asymmetry case,  the complex
structures discussed above may be preempted by the phase separation
in two uniform phases when the charge asymmetry decreases.
  \begin{figure}[h]
 \centering
 \includegraphics[height=6cm]{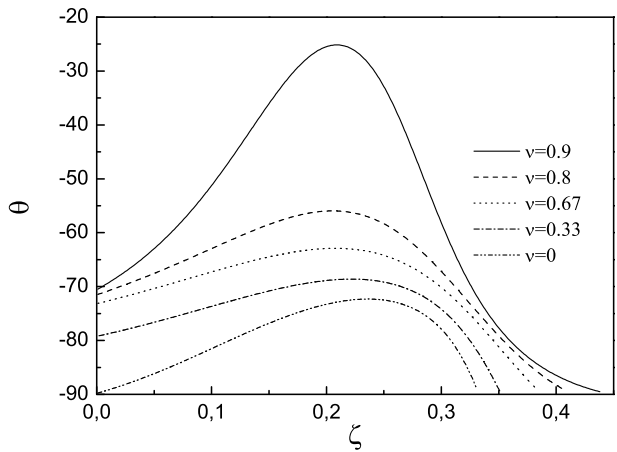}
 \caption{The variation of angle $\theta$ along the  $\lambda$-lines  presented in Fig.~11. $\theta$ is
measured in degrees.} 
\label{fig16}
 \end{figure}

\subsection{Very large asymmetry}

Let us consider the system with a very large asymmetry in size and
charge: $\delta$,$\nu\rightarrow 1$. For $\delta=0.99$ ($\lambda=200$) and $\nu=0.99$ ($Z=200$), the
 $\lambda$-line and the corresponding wave vectors  are shown in
Figs.~17 and 18. As expected,  the region bounded by the
$\lambda$-line extends to  higher temperatures in this case. The
$\lambda$-line assumes a maximum with the coordinates
$\zeta_{m}\simeq 0.09$, $ T_{m}^{*}\simeq 1.4$ and
$k_{b,m}\sigma_{\pm}\simeq 2$. Remarkably, $T_{m}^{*}$  is  five
times higher and $\zeta_{m}$  is  two times lover than for
$\delta=\nu=0.9$ ($\lambda=Z=19$). On the other hand, the above-mentioned value of
$T_{m}^{*}$ is by an order of  magnitude   smaller than that
obtained for the same model in Ref.~\cite{Ciach-Gozdz-Stell-07}.
Recall that  $\zeta_{m}$ has no  size and charge dependence in the
approximation used in the previous work. The period of the OP
oscillations $l_{b}=2\pi/k_{b}\sim \pi\sigma_{+}/2$ persists  in the
region $0<\zeta<0.25$ and increases for a higher value of the volume
fraction. The behaviour of the angle $\theta$ indicating the
direction of the dominant fluctuations is shown in Fig.~19. As is
seen,  $\theta$ changes its trend sharply from $\sim -70^{\circ}$ to
$\sim 70^{\circ}$ in the region $0\lesssim \zeta\lesssim0.05$. Then,
$\theta$ remains nearly constant ($\theta\approx 80^{\circ}$) in a
wide region of $\zeta$ ($0.05<\zeta<0.25$) and  again changes
sharply to $\sim -90^{\circ}$ keeping this value for $\zeta\lesssim
0.5$. We suggest that such a behaviour of the OP indicates the
re-entrant phase transition  observed experimentally in the
colloidal systems \cite{Yamanaka:98,Royall:2004}.
\begin{figure}[h]
\centering
\includegraphics[height=6cm]{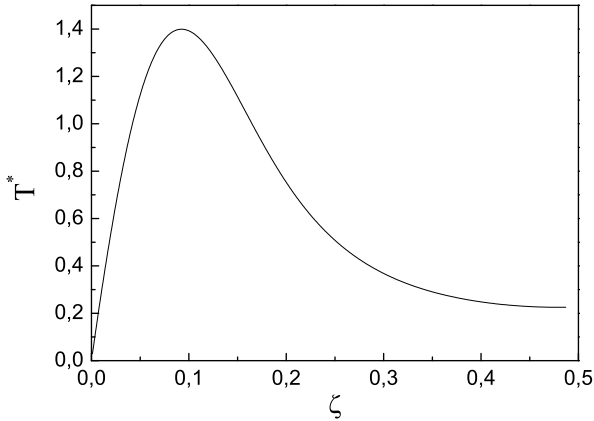}
\caption{The  $\lambda$-lines for the transition to the ordered
phase for $\delta=\nu=0.99$. Temperature $T^{*}$ and the volume
fraction of ions $\zeta$ are in dimensional reduced units defined in
Eqs.~(\ref{temp}) and (\ref{fraction}), respectively.} \label{fig17}
\end{figure}
\begin{figure}[h]
\centering
\includegraphics[height=6cm]{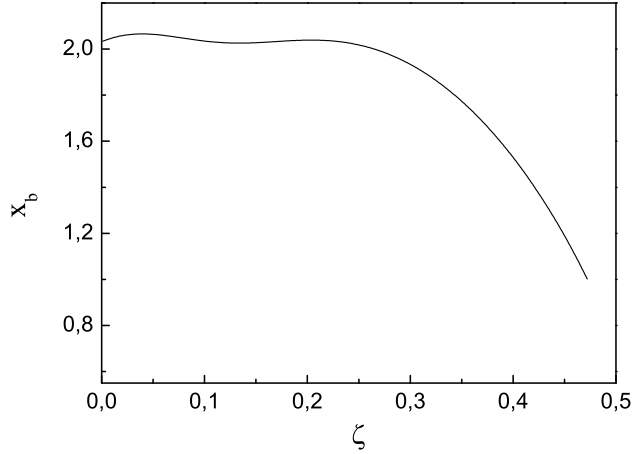}
\caption{The wave number $x_{b}=k_{b}\sigma_{\pm}$ corresponding to
the ordering of ions along the  $\lambda$-line shown in Fig.~17.
$\zeta$ is the volume fraction of ions.} \label{fig18}
\end{figure}
  \begin{figure}[h]
 \centering
 \includegraphics[height=6cm]{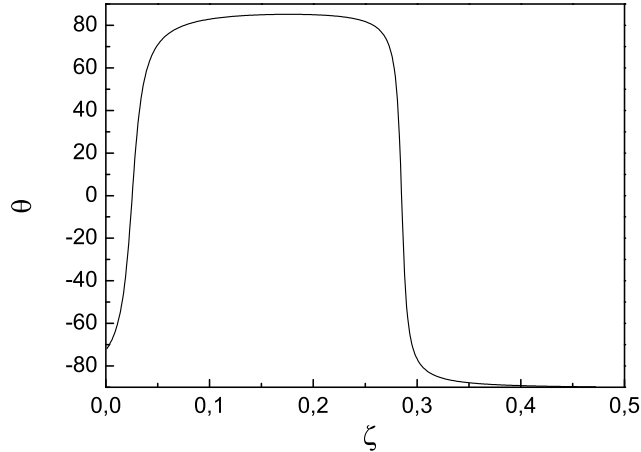}
 \caption{The variation of the angle $\theta$ along the $\lambda$-lines  shown in Fig.~17. $\theta$ is
measured in degrees.} 
\label{fig19}
 \end{figure}
\section{Discussion}

The present analysis has shown that the
size and charge asymmetry is expected to have a significant effect
on the periodic ordering in the systems with dominant Coulomb
interactions.

For the  PM with a small size asymmetry ($\lambda<2$), the trend of
the $\lambda$-lines in a wide region of the volume fractions is
qualitatively similar to that found for the RPM.  This is consistent with the behavior of  OP
demonstrating only a small deviation from the pure charge
density oscillations. In this case, we expect that the ordered phase
corresponding to an ionic crystal with a compact unit cell is formed
at lower temperatures and at higher volume fractions  when compared
to the $\lambda$-lines. In particular, such a behaviour was
confirmed experimentally for the system of oppositely charged
colloidal particles of comparable sizes (see Ref.~\cite{Leunissen}).
On the other hand,  we expect that charge-ordered living clusters
are formed in the fluid phase with the period  of the charge wave
similar to the one in the crystal.
It is worth noting that in the
case of moderate and small charge asymmetry,  the trends of the
$\lambda$-lines deviate from the RPM-like behaviour when the volume
fraction increases.

PMs with $\lambda=2$ show a crossover-type behaviour between the
regime of small size asymmetry  and the regime of moderate and large
size asymmetry.  In  qualitative agreement with the results of
Ref.~\cite{Hynnien-Panagiotopoulos:09},  our results predict the
phase transition to the ionic crystal for the system with
$\lambda=2$ and $Z=1$.

For a moderate and large size asymmetry ($2<\lambda<20$), the
$\lambda$-lines show a single maximum $T_{m}^{*}$ at  volume
fraction $\zeta_{m}$. While $T_{m}^{*}$ depends  slightly   on the
size asymmetry, $\zeta_{m}$ decreases noticeably when  the size
asymmetry increases. In addition,  $T_{m}^{*}$ decreases with the
decrease of charge asymmetry. The nonmonotonic temperature at the
$\lambda$-line indiates that  some kind of a crystalline phases can
be stable for intermediate volume fractions, and re-entrant melting
occurs at the high volume fractions (where the temperature at the
$\lambda$-line decreses). It should be  noted that both the charge
density  and the total number  density oscillate along the
$\lambda$-lines and their contributions to the OP  vary depending on
the charge asymmetry as well as on the volume fraction. Note that
for overall charge neutrality the number of negatively charged
species is $Z$ times larger than the number of positively charged
species. For a large charge asymmetry the main contribution to the
number density comes from the negatively charged species.

 The
expected phase transition to the  crystal phase of different
structure is in qualitative agreement with the simulation studies
which  predict different crystal structure in mixtures of large and
small oppositely charged spherical colloids with $\lambda\simeq 3$
\cite{Leunissen,Hynninen-06}. Remarkably, three of the predicted
structures were also observed experimentally
\cite{Leunissen,Hynninen-06}.  Our results also show   that the
large nearly neutral clusters may be formed in the PMs with a
moderate and large size asymmetry at the small volume fractions.
Such a behavior was observed experimentally in  water solutions of
the charged globular proteins \cite{stradner:04:0}. On the other
hand, the appearance of instabilities in the fluid phase against the
periodic ordering  for a moderate size and charge  asymmetry may
explain to some extent the inhomogeneous structure observed in
RTILs. In these  cases the role of non-Coulomb interactions should
be clarified in future studies. The trend of the $\lambda$-lines
found for the  large size asymmetry implies  that the periodic
ordering becomes less favorable when the charge asymmetry decreases
and the  fluid-solid phase transition may be preempted by the phase
separation in two uniform phases when the charge asymmetry
decreases.

For a very large size and charge asymmetry ($\lambda=Z=200$) the
number density fluctuations dominate except for small ranges of
$\zeta$ (Fig.~19). The period of  the OP oscillations depends very
slightly  on $\zeta$  in the region $0<\zeta<0.25$ (Fig.18). In this
case the  clusters  with diameter $\sim\pi\sigma_+/2\approx
1.5\sigma_+$ are formed. Each cluster is composed of a large
positive (negative) ion surrounded by a narrow layer of compensating
negative (positive) charges. Such clusters are associated with
pre-transitional ordering
 - they are present in the
solution when the transition to the crystal is approached. Stability
analysis alone is not sufficient for determination of the
crystalline structure. We can expect formation of the colloidal
crystal with periodic distribution of particles surrounded by a
cloud of counterions  for very  small volume fractions. It should be
noted that at the point of the $\lambda$-line corresponding to
$\zeta=0.05$ the inverse screening length is
$\kappa\sigma_{+}\approx 0.7$. The screening increases  with the
increase of $\zeta$, indicating the reduction of the long-range
repulsion along the $\lambda$-line.
 For $0.1<\zeta<0.2$ we find $\theta>80$, and the OP consists of almost pure number
density wave with the wavelength $\approx 1.5\sigma_+$, with
negligible periodic ordering of the charge density (see Fig.19). We
may thus conclude that nearly charge-neutral units composed of the
particle surrounded by a thin layer of the neutralizing counterions
are formed. Crystal formation of neutral (thus noninteracting) units
is not expected if their volume fraction is $\zeta\sim 0.1$. Thus,
 a re-entrant melting could be expected in this region. For $\zeta>0.25$ we find decreasing $\theta$ and the OP
consists of both the number and the charge waves.   The period of the OP oscillations
 increases with the increase of the volume fraction.
Such a behaviour may indicate the ion rearrangement  leading to the formation of larger clusters.

In experiments a fluid--bcc crystal--fluid phase coexistence was
found  with an increase of the colloid volume fraction
\cite{Yamanaka:98,Royall:2004}. The dilute fluid - dilute crystal -
dense fluid - dense crystal transitions, for increasing $\zeta$,
with the re-entrant melting  for  $\zeta\sim 0.1$ were also
seen~\cite{elmasri:12}. In the latter experiment room temperature in
our reduced units is $T^*\approx 0.39$. For $0.1<\zeta<0.2$ the
reduced temperature at the $\lambda$-line is much higher than
$0.39$, and further studies are necessary for comparison between our
theory and experiment. Nevertheless, we can make an observation that
crystal phases were observed for the volume fractions corresponding
to the OP consisting of both the number
 and the charge density waves, and the re-entrant melting was observed when the OP consists only of the number density wave,
 with no charge
inhomogeneity.

\section{Concluding remarks}
In this paper we have used the CVs based theory  to study the
periodic ordering in the  PMs with size and charge asymmetry. We
consider  the Gaussian approximation of the functional of the grand
partition function which, in turn, leads to the  free energy and the
direct correlation functions in the RPA. Using  analytic expressions
for  direct correlation functions in the RPA we study the effects of
the size and charge asymmetry on the instabilities of the uniform
phase with respect to the periodic ordering.

We determine the CV associated with the OP. To this
end, we diagonalize the square form of the Hamiltonian and analyze
the behaviour of the eigenvalues. This enables us to identify the
OP connected with the periodic ordering and determine
the character of the dominant fluctuations along the
 $\lambda$-lines.

We extend the study undertaken in Ref.~\cite{Ciach-Gozdz-Stell-07}
by introducing several  modifications. We  take into account the
dependence of the reference hard-sphere correlation functions on the
wave vectors and adopt the WCA regularization of the Coulomb
potential inside the hard core. While the latter  leads to a
significant decrease of the
 $\lambda$-line temperatures, the former gives rise to their increase  for  the
same regularization scheme. The combination of  the above-mentioned
modifications gives rise to the  $\lambda$-line temperatures much lower than those found
in Ref.~\cite{Ciach-Gozdz-Stell-07}.   A similar situation holds for the whole range of diameter-,
$\lambda$, and charge, $Z$, ratios of the two ionic species.  Besides, the WCA regularization scheme leads to the higher values of
the wave vectors $k_{b}$ (smaller $l_{b}=2\pi/k_{b}$) associated with the  $\lambda$-line
when compared to Ref.~\cite{Ciach-Gozdz-Stell-07}. Precise values of $k_b$ cannot be determined in a perturbative approach,
since the regularization of the Coulomb potential is not unique.

According to a typical behaviour of the  $\lambda$-lines, we
distinguish three regimes: the regime of small size asymmetry regime
($\lambda<2$), the regime of moderate and large charge asymmetry
($2<\lambda<20$) and the regime of very large size and charge
asymmetry ($\lambda=\nu=200$). As is seen, the first regime  is
found to be narrower than that considered in the previous work.
Remarkably, the qualitative dependence on the charge asymmetry
appears already  in the first regime: for a small charge asymmetry,
the  $\lambda$-lines $T^{*}$ versus $\zeta$ demonstrate the
departure from the monotonously increasing behaviour which becomes
more evident with an  increase of $\lambda$. This change in the
phase diagram  is directly related to the non-local approximation
used for the reference hard-sphere system. Unlike
Ref.~\cite{Ciach-Gozdz-Stell-07},   our results also show that the
models with $\lambda\geq 4$ and $Z<1$ (a large charge at the smaller
ion) do not undergo the instabilities with respect to the periodic
ordering in the whole region of $\zeta$. Thus, we can state
that the modifications introduced in this paper have led  to
quantitative and partly qualitative changes in the phase diagram
obtained in Ref.~\cite{Ciach-Gozdz-Stell-07}

We conclude that both the size and charge asymmetry  affect  the periodic ordering in the systems
with dominant Coulomb interactions, namely:
(1) for  $\lambda<2$,  the charge density oscillations dominate in the OP
-- the phase transition to an ionic crystal with a
compact unit cell is expected;
(2) for $\lambda>2$, both the charge density and the total number
density oscillate along the structural line  and their
contributions  to the OP  depend on $\lambda$ and $Z$;
(3) for a moderate and large $\lambda$, a crystalline phase can be
stable for intermediate volume fractions and re-entrant melting  occurs at
high volume fractions, the large nearly neutral clusters may be formed at
small volume fractions;
(4) for a large $\lambda$ and small $Z$, the fluid-crystal phase transition
can be preempted by the gas-liquid-like  phase separations;
(5) for a very large $\lambda$ and  $Z$,  the colloidal
crystal  with periodic distribution of particles surrounded by a
cloud of counterions is expected for very small volume fractions and a
re-entrant melting for the higher volume fractions.

The RPA predicts only the existence of a region in which a model
ionic fluid is unstable with respect to the OP oscillations associated
with the periodic ordering. In this context our phase diagrams
indicate the pre-transitional effects. The fluctuation effects of
the higher order than the second order should be taken into account
in order to get the information on both the  more precise location
of the  phase diagrams and  the pattern shapes which can be formed.
This will be done elsewhere.

\section*{ Acknowledgments}
Partial support by the Ukrainian-Polish joint research project under
the Agreement on Scientific Collaboration between the Polish Academy
of Sciences and the National Academy of Sciences of Ukraine for
years 2009-2011 is  gratefully acknowledged.  A part of this work  was realized within the
International PhD Projects Programme of the Foundation for Polish
Science, co-financed from European Regional Development Fund within
Innovative Economy Operational Programme ``Grants for innovation''.

\section{Appendix}
\subsection{Explicit expressions for the coefficients $t_{IJ}$}
\begin{eqnarray}
t_{NN} &=& \frac{a_{22}(k)}{\Delta}, \qquad
t_{NQ} = -\frac{a_{12}(k)}{\Delta}, \nonumber \\
t_{QN} &=& -\frac{a_{21}(k)}{\Delta}, \qquad t_{QQ} =
\frac{a_{11}(k)}{\Delta}, \label{tIJ}
\end{eqnarray}
where
\begin{eqnarray}
a_{11}(k) &=&\frac{1}{\sqrt{1+\alpha_{1}^{2}}}, \qquad
a_{12}(k)=\frac{1}{\sqrt{1+\alpha_{2}^{2}}},
\nonumber \\
a_{21}(k)& =&\frac{\alpha_{1}}{\sqrt{1+\alpha_{1}^{2}}}, \qquad
a_{22}(k)=\frac{\alpha_{2}}{\sqrt{1+\alpha_{2}^{2}}}, \label{coeff}
\end{eqnarray}
\begin{eqnarray}
 \Delta=a_{11}a_{22}-a_{12}a_{21}=\frac{|\tilde{\mathcal C}_{NQ}(k)|}{\tilde{\mathcal
 C}_{NQ}(k)},
\label{Delta}
\end{eqnarray}
\begin{eqnarray}
\alpha_{1,2}(k)=\frac{\tilde{\mathcal C}_{QQ}(k)-\tilde{\mathcal
C}_{NN}(k)\pm \sqrt{(\tilde{\mathcal C}_{NN}(k)-\tilde{\mathcal
C}_{QQ}(k))^{2}+4\tilde{\mathcal C}_{NQ}(k)^{2}}}{2\tilde{\mathcal
C}_{NQ}(k)}. \label{alpha_ij}
\end{eqnarray}

\end{document}